\documentclass[aps, pra, reprint, floatfix, superscriptaddress, showkeys, nofootinbib, fleqn]{revtex4-2}
\pdfoutput=1
\usepackage{amsmath, amsfonts, amssymb, amsthm, bm, bbm}
\usepackage[colorlinks=true, linkcolor=blue, citecolor=magenta, urlcolor=blue]{hyperref}
\usepackage{lmodern}
\usepackage[T1]{fontenc}
\usepackage[utf8]{inputenc}
\usepackage[american]{babel}
\usepackage[activate={true,nocompatibility},final,tracking=true,kerning=true,spacing=true,factor=1100]{microtype}
\microtypecontext{spacing=nonfrench}
\usepackage{orcidlink}
\usepackage{titlesec}
\usepackage{nccmath}
\usepackage{tikz-cd}
\usepackage{graphicx}
\usepackage[a4paper, lmargin=1.5cm, rmargin=1.5cm, tmargin=1.8cm, bmargin=1.8cm]{geometry}

\theoremstyle{remark}
\newtheorem{rmk}{Remark}

\newtheorem*{rmk*}{Remark}
\newtheorem*{ex*}{Example}

\newcommand{\HH}{\mathcal{H}}
\newcommand{\EE}{\mathcal{E}}

\newcommand{\tr}{\operatorname{tr}}

\newcommand{\bra}[1]{\langle #1|}
\newcommand{\ket}[1]{|#1\rangle}
\newcommand{\braket}[2]{\langle #1 | #2 \rangle}
\newcommand{\ketbra}[2]{|#1 \rangle\langle #2|}

\newcommand{\1}{\mathbbm{1}}

\newcommand{\boldr}{\boldsymbol{r}}

\newcommand{\supp}{\operatorname{supp}}

\begin{document}

\title{Systems that saturate the Margolus--Levitin quantum speed limit}
\author{Ole S{\"o}nnerborn\,\orcidlink{0000-0002-1726-4892}\,}
\email{ole.sonnerborn@kau.se}
\affiliation{Department of Mathematics and Computer Science, Karlstad University, 651 88 Karlstad, Sweden}
\affiliation{Department of Physics, Stockholm University, 106 91 Stockholm, Sweden}
\date{\today}

\begin{abstract} 
We provide a complete characterization of all finite-dimensional quantum systems that saturate the Margolus--Levitin quantum speed limit at arbitrary Uhlmann--Jozsa fidelity. Employing a purification-based approach, we prove that mixed-state saturation occurs precisely when three structural criteria are fulfilled: the state's support is confined to the sum of two energy eigenspaces (the ground level and a single excited level); each eigenvector of the state with nonzero weight is a fixed superposition of one ground- and one excited-state energy eigenvector (determined by the minimizer of the objective function identified by Giovannetti \emph{et al.}) and all such eigenvectors evolve in mutually orthogonal subspaces. These requirements impose a strict rank bound, ruling out saturation by any faithful state. For quantum bits, we derive a purity-resolved and tight Margolus--Levitin bound that reduces to the pure-state result in the limit of unit purity. Through a time-reversal argument, we further extend the dual Margolus--Levitin quantum speed limit to mixed states and establish the corresponding saturation conditions.
\end{abstract}


\maketitle

\titleformat{\section}[block]{\bfseries\large}{\Roman{section}}{0.7em}{}
\titlespacing{\section}{0em}{1.2em}{1em}
\titleformat{\subsection}[block]{\bfseries\normalsize}{\Roman{section}.\Alph{subsection}}{0.7em}{}
\titlespacing{\subsection}{0em}{1.2em}{0.5em}
\titleformat{\subsubsection}[block]{\itshape\normalsize}{\Roman{section}.\Alph{subsection}.\arabic{subsubsection}}{0.4em}{}
\titlespacing{\subsubsection}{0em}{0.5em}{0.3em}

\section{Introduction}
\label{sec: I}
\noindent
In their now-classic paper~\cite{MaLe1998}, Margolus and Levitin showed that the time $\tau$ required for a system to undergo unitary evolution between two completely distinguishable pure states is bounded from below by the ratio of $\pi/2$ to the system's expected energy above the ground-state energy (throughout, we adopt units in which $\hbar = 1$):
\begin{equation}
\tau \ge \frac{\pi/2}{E - E_0}.
\label{eq: MLQSL}
\end{equation}
This time--energy relation, commonly referred to as the Margolus--Levitin quantum speed limit, is a fundamental bound that complements the Mandelstam--Tamm time--energy uncertainty relation~\cite{MaTa1945, Fl1973, HoAlSo2022}. It plays a significant role in the conceptual foundations of quantum theory and in the quantitative analysis of dynamical processes central to emerging quantum technologies~\cite{LeTo2009, Fr2016, DeCa2017, Neetal2021, HoSo2023c}. Applications span a wide range of areas, including performance limits in quantum computation~\cite{Ll2000, Ll2002, AiDe2022}, precision metrology~\cite{GiLlMa2004, Maetal2023, Heetal2024}, bounds on information-processing rates~\cite{DeLu2010, AcDe2017, De2020}, the charging of quantum batteries~\cite{Bietal2015, Caetal2017, Lietal2025}, and quantum control and shortcuts to adiabaticity~\cite{CaDe2017}.

A natural question is how the speed limit in Eq.~\eqref{eq: MLQSL} should be modified when the requirement that the initial and final states be completely distinguishable is relaxed to allow for an arbitrary overlap $\delta$ between them. The original Margolus--Levitin argument depends crucially on the condition $\delta = 0$ and does not indicate how the inequality should be generalized for $0 < \delta < 1$. Shortly after the Margolus--Levitin result appeared, however, Giovannetti \emph{et al.}~\cite{GiLlMa2003} suggested the following extension of the Margolus--Levitin quantum speed limit:
\begin{equation}
\tau \geq \frac{\alpha(\delta)}{E - E_0},
\quad
\alpha(\delta)
= \min_{z^2 \leq \delta}(1+z)\arcsin\sqrt{\frac{1-\delta}{1-z^2}}.
\label{eq: GQSL}
\end{equation}
The inequality in Eq.~\eqref{eq: GQSL} can be derived directly for qubits, and, supported by numerical evidence, Giovannetti \emph{et al.} conjectured that it should hold for arbitrary finite-dimensional systems as well. Remarkably, a complete analytical proof did not appear until two decades later~\cite{HoSo2023a}. This delay can be attributed in part to the fact that the Margolus--Levitin speed limit possesses a geometric character that differs fundamentally from that of other quantum speed limits, as discussed in Ref.~\cite{HoSo2023a}.

More realistic descriptions of quantum systems must allow for statistical ensembles of pure states, or mixed states. Assuming the validity of Eq.~\eqref{eq: GQSL} for pure states, Giovannetti \emph{et al.} further showed that the same bound continues to hold for mixed states provided that the overlap is replaced by the Uhlmann--Jozsa fidelity~\cite{Uh1976, Jo1994}. Because the structure of their argument plays an important role in the developments presented here, we include in Sec.~\ref{sec: II} a slightly adapted version of their original derivation.

In this work, we identify all systems that saturate the Margolus--Levitin quantum speed limit, that is, those whose dynamics achieve equality in Eq.~\eqref{eq: GQSL}. Specifically, for any finite-dimensional system evolving under a time-independent Hamiltonian, we derive necessary and sufficient conditions on the initial state ensuring that it evolves into a state of prescribed fidelity in a time saturating the Margolus--Levitin bound. While Ref.~\cite{HoSo2023a} provides a complete characterization of the pure states attaining the Margolus--Levitin bound, we extend this characterization to mixed states (Secs.~\ref{sec: III} and~\ref{sec: IV}), thereby broadening the class of states for which the bound is tight.

From this characterization, it follows that any state saturating the Margolus--Levitin quantum speed limit must have a highly constrained rank. In particular, its support is confined to the sum of two energy eigenspaces, and its dimension cannot exceed the smaller of the multiplicities of those eigenvalues. Consequently, faithful states can never saturate the Margolus--Levitin bound. This observation is made explicit in Sec.~\ref{sec: V}, where we present a refined Margolus--Levitin quantum speed limit applicable to mixed---and therefore faithful---quantum bits.

In Sec.~\ref{sec: V}, we also extend the dual Margolus--Levitin quantum speed limit derived by Ness \emph{et al.}~\cite{NeAlSa2022} to mixed states and characterize the states for which this bound is saturated. This characterization follows directly from that obtained for systems saturating the standard Margolus--Levitin bound, since the dual speed limit can be recovered from the original via a time-reversal argument.

\section{The Margolus--Levitin speed limit}
\label{sec: II}
\noindent 
Consider a system in a state $\rho$ undergoing unitary evolution generated by a Hamiltonian $H$. Let $E = \tr(\rho H)$ denote the system's expected energy, and let $E_0$ be the ground-state energy, defined as the smallest populated eigenvalue of $H$ (see Remark~\ref{rmk: 2} below). The time $\tau$ required for the system to evolve into a state having Uhlmann--Jozsa fidelity $\delta$ with the initial state is then bounded from below by the Margolus--Levitin quantum speed limit:
\begin{equation}
    \tau \geq \frac{\alpha(\delta)}{E - E_0}, \quad \alpha(\delta)=\min f_\delta(z),
    \label{eq: the QSL}
\end{equation}
where the objective function is
\begin{equation}
f_\delta(z)
= (1+z)\arcsin\sqrt{\frac{1-\delta}{1-z^2}},\quad -\sqrt{\delta}\leq z\leq \sqrt{\delta}.
\label{eq: objective function}
\end{equation}

The Uhlmann--Jozsa fidelity (hereafter simply \emph{fidelity}) is a measure of distinguishability between quantum states and is defined in terms of the trace norm as
\begin{equation}
	F(\rho_1, \rho_2)
	= \| \sqrt{\rho_1} \sqrt{\rho_2} \|_{\tr}^2.
\end{equation}
The trace norm of an operator is given by $\|A\|_{\tr} = \tr|A|$, where $|A| = \sqrt{A^\dagger A}$. For pure states, the fidelity reduces to their overlap: if $\rho_1=\ketbra{\psi_1}{\psi_1}$ and $\rho_2=\ketbra{\psi_2}{\psi_2}$, then 
\begin{equation}
    F(\rho_1, \rho_2) = \tr(\rho_1 \rho_2) = |\braket{\psi_1}{\psi_2}|^2.
\end{equation}

That the inequality in Eq.~\eqref{eq: the QSL} holds for systems in pure states was established in Ref.~\cite{HoSo2023a}. Following Giovannetti \emph{et al.}~\cite{GiLlMa2003}, we show that this result implies its validity for systems in mixed states as well.

\begin{rmk}
An energy level (an eigenvalue of $H$) is \emph{populated} if the support of the state is not orthogonal to the corresponding eigenspace; equivalently, the eigenvalue occurs with nonzero probability in an energy measurement. For a time-independent Hamiltonian, these probabilities are conserved, so the state's support remains within the sum of the populated eigenspaces.

In the literature on the Margolus--Levitin quantum speed limit, $E_0$ is typically defined as the smallest eigenvalue of $H$. If this eigenvalue is unpopulated, the smallest populated eigenvalue can be used instead, yielding a strictly larger lower bound on the evolution time. In this work, we take $E_0$ to be the smallest populated eigenvalue.
\label{rmk: 2}
\end{rmk}

\begin{rmk}
The Margolus--Levitin speed limit does not, in general, extend to systems with time-dependent Hamiltonians. As shown in Ref.~\cite{HoSo2023b}, no canonical extension of the Margolus--Levitin bound exists in this setting.
\label{rmk: 3}
\end{rmk}

\subsection{Purification and Uhlmann's theorem}
\label{sec: Purification and Uhlmann theorem}
\noindent 
Before demonstrating that the Margolus--Levitin speed limit extends to systems in mixed states, we first introduce the notion of purification and recall Uhlmann's theorem.

For every mixed state $\rho$ acting on a Hilbert space $\HH$, there exists a unit vector $\ket{w} \in \HH \otimes \HH$ such that $\rho$ is recovered by taking the partial trace of $\ketbra{w}{w}$ over the second copy of $\HH$: $\rho = \tr_2 \ketbra{w}{w}$; see Refs.~\cite{Jo1994, NiCh2010}. This vector is called a \emph{purification} of $\rho$.

The Hamiltonians $H$ and $H \otimes \1$ share the same eigenvalue spectrum, up to multiplicities, and an eigenvalue of $H$ is populated by $\rho$ if and only if the corresponding eigenvalue of $H \otimes \1$ is populated by $\ketbra{w}{w}$. (Accordingly, $E_0$ denotes the smallest populated eigenvalue for both $\rho$ and $\ketbra{w}{w}$.) Moreover, $\rho$ and $\ketbra{w}{w}$ have the same expected energy:
\begin{equation}
	\tr(H\rho) = \bra{w} H \otimes \1 \ket{w}.
\end{equation}

Any other purification of $\rho$ has the form $\1 \otimes U \ket{w}$ for some unitary operator $U$ acting on $\HH$. Uhlmann's theorem~\cite{Uh1976} asserts that the fidelity between two mixed states equals the maximal possible overlap between their purifications. Specifically, if $\ket{w_1}$ and $\ket{w_2}$ are purifications of $\rho_1$ and $\rho_2$, respectively, then
\begin{equation}
	F(\rho_1, \rho_2) = \max_U |\bra{w_1}\, \1 \otimes U \ket{w_2}|^2,
	\label{eq: Uhlmann theorem}
\end{equation}
where the maximization is taken over all unitary operators $U$ on $\HH$. A proof of Uhlmann's theorem can be found in Ref.~\cite{Uh1976} and in Refs.~\cite{Jo1994} and~\cite{NiCh2010}.

\subsection{The Margolus--Levitin quantum speed limit for systems in mixed states}
\label{sec: Derivation of Margolus--Levitin quantum speed limit}
\noindent
Consider a system with Hamiltonian $H$ prepared in a state $\rho$. Its time evolution is given by
\begin{equation}
	\rho_t = U_t \rho U_t^\dagger, 
	\quad 
	U_t = e^{-i t H}.
	\label{eq: evolution of state}
\end{equation}
Let $\ket{w}$ be a purification of $\rho$.
The corresponding time-evolved purification is defined as
\begin{equation}
	\ket{w_t} = U_t \otimes \1 \ket{w},
	\label{eq: evolution of purification}
\end{equation}
which is the solution of the Schr\"odinger equation generated by $H \otimes \1$ starting from $\ket{w}$.

The vector $\ket{w_\tau}$ is a purification of the final state $\rho_\tau$, and, by Uhlmann's theorem, the overlap $|\braket{w}{w_\tau}|^2$ is upper bounded by the fidelity  $F(\rho, \rho_\tau)$.  As illustrated in Fig.~\ref{fig: alpha} and proven in Appendix~\ref{app: A}, the function $\alpha$ is monotonically decreasing. Consequently, $\alpha (|\braket{w}{w_\tau}|^2)$ is lower bounded by $\alpha( F(\rho, \rho_\tau))$. Using the Margolus--Levitin bound for pure states, we therefore obtain
\begin{equation}
	\tau 
	\geq \frac{\alpha(|\braket{w}{w_\tau}|^2)}{E - E_0}
	\geq \frac{\alpha(F(\rho, \rho_\tau))}{E - E_0}.
	\label{eq: inequality by inequality}
\end{equation}
This establishes that the Margolus--Levitin quantum speed limit holds for systems in mixed states as well.

\begin{figure}[t]
\centering
\includegraphics[width=\linewidth]{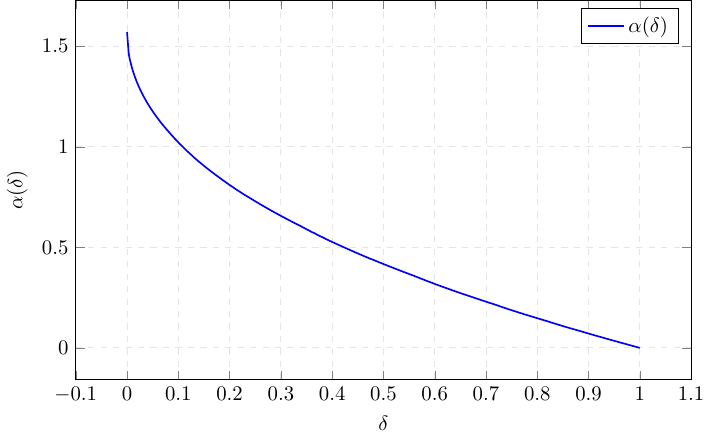}
\caption{Plot of the function $\alpha$, which appears in the numerator of the Margolus--Levitin bound, shown as a function of the fidelity $\delta$ between the initial and final states. As evident from the plot, the function is strictly decreasing. A proof of this property is provided in Appendix~\ref{app: A}.}
\label{fig: alpha}  
\end{figure}

\section{Systems saturating the Margolus--Levitin quantum speed limit}
\label{sec: III}
\noindent
In this section, we derive necessary and sufficient conditions on a system's state guaranteeing saturation of the Margolus--Levitin quantum speed limit. Reference~\cite{HoSo2023a} provides a complete characterization of the \emph{pure} states for which the system attains the bound. We begin by recalling that result.

\subsection{Systems in pure states that saturate the Margolus--Levitin quantum speed limit}
\label{sec: Pure state systems saturating Margolus--Levitin quantum speed limit}
\noindent
Suppose the initial state of the system is represented by the unit vector $\ket{\psi}$. This state evolves into one whose overlap with $\ketbra{\psi}{\psi}$ equals $\delta$ in a time that saturates the Margolus--Levitin bound if and only if $\ket{\psi}$ can be written as a superposition of the form
\begin{equation}
    \ket{\psi}
    = \sqrt{\frac{1 - z_\delta}{2}} \ket{E_0}
    + \sqrt{\frac{1 + z_\delta}{2}} \ket{E_m},
    \label{eq: the pure state}
\end{equation}
where $\ket{E_0}$ is a normalized ground-state vector, $\ket{E_m}$ is a normalized energy eigenvector corresponding to a higher eigenvalue $E_m$, and $z_\delta$ is the unique value of $z$ at which the objective function $f_\delta(z)$, defined in Eq.~\eqref{eq: objective function}, attains its minimum. The existence and uniqueness of this value are established in Appendix~\ref{app: B}. Figure~\ref{fig: f} plots $f_\delta(z)$ for several values of $\delta$, with the associated minimizers indicated along the horizontal axis.

\begin{figure}[t]
    \centering
    \includegraphics[width=0.98\linewidth]{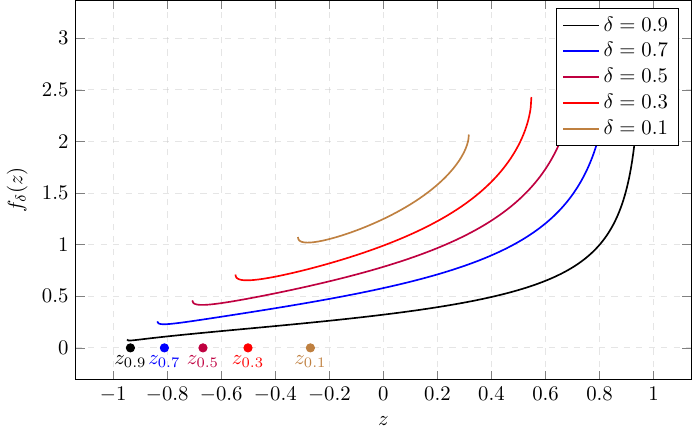}
    \caption{Plots of the objective functions corresponding to the fidelities $\delta = 0.9, 0.7, 0.5, 0.3, 0.1$. The $z$-coordinates of their minima are indicated along the horizontal axis. A proof that the objective function for a given fidelity possesses a unique minimum is provided in Appendix~\ref{app: B}.}
    \label{fig: f}
\end{figure}

This characterization follows from the observation that the system must effectively behave as a qubit, that is, evolve within the span of two eigenvectors of the Hamiltonian. This was shown in Ref.~\cite{HoSo2023a}. For pure qubits, it is then straightforward to determine which states saturate the Margolus--Levitin bound. An explicit proof that $\ket{\psi}$ can be written in the form of Eq.~\eqref{eq: the pure state} is provided in Sec.~\ref{sec: V}, which analyzes qubits in detail.

\subsection{Systems in mixed states that saturate the Margolus--Levitin quantum speed limit}
\label{sec: Mixed state systems saturating Margolus--Levitin quantum speed limit}

\noindent
If a system with Hamiltonian $H$ evolves between two states with fidelity $\delta$ in a time $\tau$ that saturates the Margolus--Levitin quantum speed limit~\eqref{eq: the QSL}, then the following statements hold for its initial state $\rho$:
\begin{itemize}
    \item[(i)]
    The support of $\rho$ is contained within the sum of two eigenspaces of $H$.
\end{itemize}

Since the populations of the energy levels are conserved, the system effectively reduces to a two-level system, possibly with degenerate levels. The lower of the two populated energies is $E_0$, and we denote the other energy by $E_m$.

\begin{itemize}
\item[(ii)]
    Every normalized eigenvector of $\rho$ with a nonzero eigenvalue can be written in the form specified in Eq.~\eqref{eq: the pure state}.
\end{itemize}

This property implies---and is in fact equivalent to---the statement that all eigenstates of $\rho$ with nonzero eigenvalues share the same expected energy, and that each evolves between states with overlap $\delta$ in a time $\tau$ that saturates the Margolus--Levitin bound. Giovannetti \emph{et al.}~\cite{GiLlMa2003} noted this as well, assuming the validity of the Margolus--Levitin quantum speed limit for pure states.  Their argument, however, relied on numerical evidence concerning the function $\alpha$. In contrast, we obtain the same result here using only analytically established properties.

\begin{itemize}
    \item[(iii)]
    Any two orthogonal eigenvectors of $\rho$ with nonzero eigenvalues evolve within orthogonal two-dimensional subspaces of the Hilbert space.
\end{itemize}

This property implies that the rank of the state cannot exceed the smaller of the multiplicities of $E_0$ and $E_m$. Consequently, a faithful state can never saturate the Margolus--Levitin bound. In Sec.~\ref{sec: V}, we refine the Margolus--Levitin speed limit for qubits in a manner that remains tight irrespective of the state's purity.

\subsubsection*{Equivalent description of systems that saturate the Margolus--Levitin quantum speed limit}
\noindent
As we show below, conditions (i)--(iii) are jointly necessary and sufficient for a system to saturate the Margolus--Levitin quantum speed limit. They admit an equivalent formulation as follows. In a spectral decomposition
\begin{equation}
	\rho = \sum_{j=1}^r p_j \ketbra{\psi_j}{\psi_j},
	\label{eq: spectral decomposition}
\end{equation}
where $r$ is the rank of $\rho$ and thus all $p_j>0$, each eigenvector admits a representation of the form
\begin{equation}
    \ket{\psi_j}
    = \sqrt{\frac{1 - z_\delta}{2}} \ket{E_0^{j}}
    + \sqrt{\frac{1 + z_\delta}{2}} \ket{E_m^{j}},
    \label{eq: representation}
\end{equation}
where $\ket{E_0^{j}}$ and $\ket{E_m^{j}}$ are normalized eigenvectors of $H$ associated with the eigenvalues $E_0$ and $E_m$, respectively. These energy eigenvectors are pairwise orthogonal,
\begin{equation}
	\braket{E_0^{j}}{E_0^{k}} = \braket{E_m^{j}}{E_m^{k}} = \delta_{jk},
	\quad
	\braket{E_0^{j}}{E_m^{k}} = 0,
	\label{eq: orthogonal}
\end{equation}
which means that the eigenvectors $\ket{\psi_j}$ evolve within mutually orthogonal two-dimensional subspaces of the Hilbert space. All pure states $\ketbra{\psi_j}{\psi_j}$ have the same expected energy, namely
\begin{equation}
E = \bigg( \frac{1 - z_\delta}{2} \bigg) E_0
+ \bigg( \frac{1 + z_\delta}{2} \bigg) E_m,
\label{eq: the energy}
\end{equation}
and each evolves in such a way that it saturates the Margolus--Levitin bound with overlap $\delta$ at time $\tau$.

\section{Proof that properties (i)--(iii) are necessary and sufficient}
\label{sec: IV}
\noindent
In establishing properties (i)--(iii), we rely on certain properties of the function $\alpha$ and of the trace norm. We begin by stating these properties.

\subsection{Strict monotonic decay of $\alpha$}
\label{sec: Properties of alpha}
\noindent
The function $\alpha$ appearing in the numerator of the Margolus--Levitin quantum speed limit is strictly decreasing: if $\delta_1 < \delta_2$, then $\alpha(\delta_1) > \alpha(\delta_2)$. This behavior is apparent from Fig.~\ref{fig: alpha}, and a proof is given in Appendix~\ref{app: A}.

\subsection{Properties of the trace norm}
\label{sec: properties of the trace norm}
\noindent
The trace norm is unitarily invariant, meaning that $\tr|UAV| = \tr|A|$ for all unitary operators $U$ and $V$. Moreover, $|\tr A| \leq \tr|A|$, with equality if and only if $A$ differs from $|A|$ only by a unimodular factor:
\begin{equation}
    |\tr A| = \tr|A| 
    \iff A = e^{i\theta}|A| \text{ for some } \theta \in \mathbb{R}.
    \label{eq: trace identity}
\end{equation}
A proof of this equivalence is given in Appendix~\ref{app: C}.

\subsection{Properties (i)--(iii) are necessary}
\label{sec: Derivation of necessary conditions}
\noindent
Consider a system with Hamiltonian $H$ that evolves between two states whose fidelity is $\delta$ in a time $\tau$ that saturates the Margolus--Levitin quantum speed limit~\eqref{eq: the QSL}. Let $\rho$ be the initial state, $E = \tr(\rho H)$ its expected energy, and $E_0$ the lowest populated energy level. Let $\rho_t$ be the time-evolved state as defined in Eq.~\eqref{eq: evolution of state}.

Choose an arbitrary purification $\ket{w}$ of $\rho$ and let it evolve in time as in Eq.~\eqref{eq: evolution of purification}. The operators $H\otimes\1$ and $H$ are isospectral (up to multiplicities), and $E_0$ is the smallest eigenvalue of $H\otimes\1$ that is populated by $\ketbra{w}{w}$. Moreover, $\bra{w}H\otimes\1\ket{w} = E$.

Since equality is assumed to hold throughout Eq.~\eqref{eq: inequality by inequality}, and since $\alpha$ is strictly decreasing and therefore injective, it follows that
\begin{equation}
    |\braket{w}{w_\tau}|^2 = F(\rho, \rho_\tau).
    \label{eq: equal}
\end{equation}
The first equality in Eq.~\eqref{eq: inequality by inequality} then implies that the Margolus--Levitin bound for pure states is saturated by $\ketbra{w}{w}$. From the characterization in Sec.~\ref{sec: III}, $\ket{w}$ must therefore lie in the sum of the eigenspaces of $H\otimes\1$ associated with $E_0$ and some higher eigenvalue $E_m$. These eigenspaces are $\EE_0\otimes\HH$ and $\EE_m\otimes\HH$, where $\EE_0$ and $\EE_m$ are the eigenspaces of $H$ corresponding to $E_0$ and $E_m$. Since
\begin{equation}
	\ket{w} \in (\EE_0 \otimes \HH) \oplus (\EE_m \otimes \HH)
	= (\EE_0 \oplus \EE_m) \otimes \HH,
\end{equation}
the support of $\rho$ is contained in the sum of $\EE_0$ and $\EE_m$:
\begin{equation}
	\supp\rho = \supp\tr_2\ketbra{w}{w} \subset \EE_0 \oplus \EE_m.
\end{equation}
This establishes condition~(i).

Properties (ii) and (iii) follow from consequences of Eq.~\eqref{eq: equal}. According to this equation, the square root of the overlap between the purifications $\ket{w}$ and $\ket{w_\tau}$ coincides with the square root of the fidelity between $\rho$ and $\rho_\tau$. Both admit the following representations: \begin{subequations}
\begin{align}
    &\hspace{-5pt}|\braket{w}{w_\tau}|
	= |\tr(\rho\, U_\tau)|
	= |\tr(\sqrt{\rho}\,U_\tau\sqrt{\rho})|,
	\label{eq: pure overlap} \\
    &\hspace{-5pt}\sqrt{F(\rho,\rho_\tau)}
	= \tr|\sqrt{\rho}\,U_\tau \sqrt{\rho}\,U_\tau^\dagger|
	= \tr|\sqrt{\rho}\,U_\tau \sqrt{\rho}\,|,
	\label{eq: mixed overlap}
\end{align}
\end{subequations}
where in the last step of Eq.~\eqref{eq: mixed overlap} we have used the unitary invariance of the trace norm. Since the right-hand sides of Eqs.~\eqref{eq: pure overlap} and~\eqref{eq: mixed overlap} coincide, Eq.~\eqref{eq: trace identity} implies that
\begin{equation}
	\sqrt{\rho}\,U_\tau\sqrt{\rho}
	= e^{i\theta}\,|\sqrt{\rho}\,U_\tau \sqrt{\rho}\,|
	\text{ for some } \theta \in \mathbb{R}.
	\label{eq: trace identity II}
\end{equation}

In Appendix~\ref{app: D}, we show that the condition in Eq.~\eqref{eq: trace identity II} implies that the compression of $U_\tau$ to the support of $\rho$ is proportional to the orthogonal projection onto that support, and that the proportionality constant has modulus $\sqrt{\delta}$. Thus, if $P$ denotes the orthogonal projection onto the support of $\rho$, Eq.~\eqref{eq: trace identity II} yields
\begin{equation}
	P\,U_\tau\,P = \sqrt{\delta}\,e^{i\theta}P.
	\label{eq: proportional II}
\end{equation}

To establish property~(ii), let $\ket{\psi}$ be any normalized eigenvector of $\rho$ with a nonzero eigenvalue $p$. Equation~\eqref{eq: proportional II} then implies that $\ketbra{\psi}{\psi}$ evolves into a state whose overlap with $\ketbra{\psi}{\psi}$ equals $\delta$ at time~$\tau$:
\begin{equation}
	|\bra{\psi} U_\tau \ket{\psi}|^2 = \delta.
	\label{eq: delta overlap}
\end{equation}

Let $\ket{\psi_1} = \ket{\psi}$ and $p_1 = p$, and expand $\rho$ in a spectral decomposition as in Eq.~\eqref{eq: spectral decomposition}, where $r$ is the rank of $\rho$, so that all $p_j > 0$ and each $\ket{\psi_j}$ lies in the support of $\rho$. We now show that all states $\ketbra{\psi_j}{\psi_j}$ have the same expected energy. Since each of these states is supported in $\EE_0 \oplus \EE_m$ and is nonstationary, it necessarily populates the ground-energy level. Applying the Margolus--Levitin bound together with Eq.~\eqref{eq: delta overlap} yields, for every $\ket{\psi_j}$,
\begin{equation}
	\tau\big( \bra{\psi_j} H \ket{\psi_j} - E_0 \big)
	\geq \alpha \big( |\bra{\psi_j} U_\tau \ket{\psi_j}|^2 \big)
	= \alpha(\delta),
\end{equation}
and therefore
\begin{equation}
	\bra{\psi_j} H \ket{\psi_j}
	\geq \frac{\alpha(\delta)}{\tau} + E_0 
	= E.
	\label{eq: energy identity}
\end{equation}
Equality with $E$ follows from the assumption that the Margolus--Levitin quantum speed limit is saturated. Because $E = \sum_{j=1}^r p_j \bra{\psi_j} H \ket{\psi_j}$, each term in this sum must individually satisfy
\begin{equation}
	\bra{\psi_j} H \ket{\psi_j} = E.
	\label{eq: energy equality}
\end{equation}

Equations~\eqref{eq: delta overlap} and~\eqref{eq: energy equality} together show that every eigenvector---and in particular $\ket{\psi}$---represents a state whose unitary evolution saturates the Margolus--Levitin speed limit. It follows that $\ket{\psi}$ can be represented in the form given in Eq.~\eqref{eq: the pure state}, in accordance with the characterization of pure states that achieve equality in the Margolus--Levitin bound. This establishes condition~(ii).

To establish~(iii), we represent each eigenvector $\ket{\psi_j}$ as in Eq.~\eqref{eq: representation}. We then show that the associated energy eigenvectors are mutually orthogonal, as expressed in Eq.~\eqref{eq: orthogonal}, from which condition~(iii) follows.

Let $P_0$ and $P_m$ be the orthogonal projections onto the eigenspaces $\EE_0$ and $\EE_m$, respectively. Since the support of $\rho$ is contained in $\EE_0 \oplus \EE_m$, Eq.~\eqref{eq: proportional II} implies that
\begin{equation}
	e^{-i\tau E_0} P P_0 P + e^{-i\tau E_m} P P_m P = \sqrt{\delta}\,e^{i\theta} P.
\end{equation}
Together with $PP_0P+PP_mP=P$, this shows that the compressions of the eigenspace projectors to the support of $\rho$ are each proportional to $P$:
\begin{subequations}
\begin{alignat}{2}
    P P_0 P &= q_0 P,\quad && q_0 = 
    \frac{\sqrt{\delta}\,e^{i\theta} 
    	- e^{-i\tau E_m}}{e^{-i\tau E_0} - e^{-i\tau E_m}},
    \label{eq: one-a} \\
    P P_m P &= q_m P,\quad && q_m =
    \frac{\sqrt{\delta}\,e^{i\theta} 
    	- e^{-i\tau E_0}}{e^{-i\tau E_m} - e^{-i\tau E_0}}.
    \label{eq: one-b}
\end{alignat}
\end{subequations}
Combining these relations with the representation in Eq.~\eqref{eq: representation} yields, for all indices 
$j,k$,
\begin{subequations}
\begin{align}
	q_0\delta_{jk} &= \bra{\psi_k} P_0 \ket{\psi_j}
	= \frac{1}{2}(1 - z_\delta) \braket{E_0^{j}}{E_0^{k}}, \\
	q_m\delta_{jk} &= \bra{\psi_k} P_m \ket{\psi_j}
	= \frac{1}{2}(1 + z_\delta) \braket{E_m^{j}}{E_m^{k}}.
\end{align}
\end{subequations}
Since $z_\delta\ne \pm 1$, these identities show that the conditions in Eq.~\eqref{eq: orthogonal} are satisfied and hence that the eigenvectors $\ket{\psi_j}$ evolve within mutually orthogonal two-dimensional subspaces.

\subsection{Properties (i)--(iii) are sufficient}
\label{sec: sufficient conditions}
\noindent
In the preceding section, we showed that conditions~(i)--(iii) are necessary, in the sense that they must hold whenever the Margolus--Levitin speed limit is saturated. Here, we demonstrate that these conditions are also sufficient.

Assume that a system with Hamiltonian $H$ is in a state $\rho$ whose support is contained in the sum of the eigenspaces associated with two energy eigenvalues $E_0$ and $E_m$, with $E_0 < E_m$. Further assume that every eigenvector of $\rho$ with a nonzero eigenvalue admits a representation of the form in Eq.~\eqref{eq: the pure state}, so that the corresponding pure state evolves into a state with overlap $\delta$ after a time $\tau$ that saturates the Margolus--Levitin bound. Finally, let $\rho$ have a spectral decomposition of the form in Eq.~\eqref{eq: spectral decomposition}, and assume that the eigenvectors in this decomposition evolve within mutually orthogonal subspaces. Under these conditions,
\begin{equation}
\begin{split}
    \sqrt{\rho}\, U_\tau \sqrt{\rho} 
    &= \sum_{j=1}^r \sum_{k=1}^r \sqrt{p_j p_k}\, 
       \ket{\psi_j} \bra{\psi_j} U_\tau \ket{\psi_k} \bra{\psi_k} \\
    &= \sum_{j=1}^r p_j \ket{\psi_j} \bra{\psi_j} U_\tau \ket{\psi_j} \bra{\psi_j} \\
    &= \frac{1}{2}\big( (1 - z_\delta)e^{-i\tau E_0}
       + (1 + z_\delta)e^{-i\tau E_m} \big) \rho.
\end{split}
\end{equation}
Thus, $\sqrt{\rho} U_\tau \sqrt{\rho}$ is proportional to $\rho$, which is positive semidefinite. Hence Eq.~\eqref{eq: trace identity II} holds, and therefore so does Eq.~\eqref{eq: equal}. From Eq.~\eqref{eq: equal}, we obtain
\begin{equation}
    \sqrt{F(\rho,\rho_\tau)}
    = \sum_{j=1}^r p_j\, |\braket{\psi_j}{\psi_{j;\tau}}|
    = \sqrt{\delta}.
\end{equation}
Accordingly, the system evolves between two states with fidelity $\delta$ in time~$\tau$. The smallest populated eigenvalue of $H$ is $E_0$, and the expected energy is
\begin{equation}
    \tr(\rho H)
    = \sum_{j=1}^r p_j \bra{\psi_j} H \ket{\psi_j}
    = E.
\end{equation}
Thus,
\begin{equation}
    \tau
    = \frac{\alpha(\delta)}{E - E_0}
    = \frac{\alpha(F(\rho,\rho_\tau))}{\tr(\rho H) - E_0}.
\end{equation}
This shows that the Margolus--Levitin quantum speed limit is saturated whenever properties~(i)--(iii) are satisfied.

\section{Discussion}
\label{sec: V}
\noindent
The Margolus--Levitin quantum speed limit \eqref{eq: the QSL} applies to both pure and mixed states, and in this work we have identified the precise circumstances under which it can be saturated. Specifically, if a system evolves unitarily from an initial state $\rho$ to a state with fidelity $\delta$ in a time $\tau$ that attains the Margolus--Levitin bound, then $\rho$ must be a convex combination of pure states that populate the same two energy eigenvalues, have identical expected energy, and individually saturate the bound with overlap $\delta$ at time $\tau$. Moreover, the supports of these pure states must evolve within mutually orthogonal subspaces.

An immediate consequence of these structural constraints is that the rank of $\rho$ cannot exceed the degeneracy of the ground-state energy. This requirement rules out faithful states. Below we introduce a modified Margolus--Levitin bound for qubits that permits faithful states to saturate the resulting speed limit as well.

This modification naturally raises the question of whether Uhlmann--Jozsa fidelity is the most appropriate extension of the pure-state overlap to mixed states in this setting. While fidelity is a standard and operationally meaningful choice, it is by no means unique; several alternative generalizations of overlap exist and may yield distinct or sharper speed limits. Investigating such alternatives will be the subject of future work.

Ness \emph{et al.}~\cite{NeAlSa2022} derived a dual version of the Margolus--Levitin quantum speed limit, which was later extended to arbitrary overlap in Ref.~\cite{HoSo2023a}. In the last section below we extend this dual bound to mixed states, using Uhlmann--Jozsa fidelity as the distinguishability measure, and characterize all mixed-state systems that achieve equality in this dual quantum speed limit.

\subsection{Quantum bits}
\label{sec: Quantum bits}
\noindent
The rank of any state that saturates the Margolus--Levitin speed limit cannot exceed the smaller degeneracy of the two energy levels it populates. As a consequence, no mixed qubit can satisfy the Margolus--Levitin bound. In this section, we derive a qubit-specific variant of the Margolus--Levitin bound that overcomes this limitation. The derivation will also clarify the structural origin of Eq.~\eqref{eq: the pure state}.

Consider a qubit governed by the Hamiltonian
\begin{equation}
    H = E_0\ketbra{E_0}{E_0} + E_m\ketbra{E_m}{E_m}, \quad E_0 < E_m.
\end{equation}
We represent the state $\rho$ by its Bloch vector $\boldr = (x,y,z)$ in three-dimensional Euclidean space, defined by
\begin{subequations}
\begin{align}
    x &= \bra{E_m}\rho\ket{E_0} + \bra{E_0}\rho\ket{E_m}, \label{x}\\
    y &= i\!\left(\bra{E_m}\rho\ket{E_0} - \bra{E_0}\rho\ket{E_m}\right), \label{y}\\
    z &= \bra{E_m}\rho\ket{E_m} - \bra{E_0}\rho\ket{E_0}. \label{z}
\end{align}
\end{subequations}
The $z$-component of the Bloch vector is determined by the expected energy $E = \tr(\rho H)$ through
\begin{equation}
    z = \frac{2(E - E_0)}{E_m - E_0} - 1,
    \label{eq: the z-coordinate}
\end{equation}
and the length of the Bloch vector is fixed by the state's purity $\wp = \tr(\rho^2)$, which remains constant under unitary evolution:
\begin{equation}
    |\boldr|^2 = 2\wp - 1.
\end{equation}

\begin{figure}[t]
    \centering
    \includegraphics[width=0.7\linewidth]{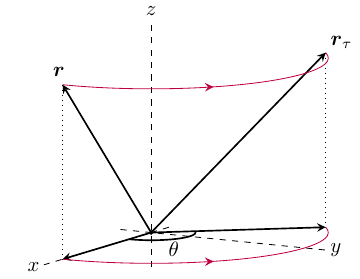}
    \caption{During the evolution, the Bloch vector precesses about the $z$-axis (vertical dashed line) with fixed inclination and angular speed $\omega$. The projections of the initial Bloch vector $\boldr$ and final Bloch vector $\boldr_\tau$ onto the $xy$-plane subtend an angle $\theta$, yielding an evolution time $\tau=\theta/\omega$.}
    \label{fig: Bloch rotation}
\end{figure}

The dynamics generated by $H$ cause the Bloch vector to rotate about the $z$-axis with constant inclination and angular speed $\omega = E_m - E_0$. Let $\boldr_\tau$ denote the Bloch vector of the state at time $\tau$. During the evolution, the orthogonal projection of the Bloch vector onto the $xy$-plane rotates through an angle
\begin{equation}
	\theta 
	\geq \arccos\!\bigg( \frac{\boldr \cdot \boldr_\tau - z^2}{|\boldr|^2 - z^2} \bigg)
	= \arccos\!\bigg( \frac{\boldr \cdot \boldr_\tau - z^2}{2\wp - 1 - z^2} \bigg).
    \label{eq: the angle}
\end{equation}
An illustration is provided in Fig.~\ref{fig: Bloch rotation}. The corresponding evolution time is therefore
\begin{equation}
	\tau 
	= \frac{\theta}{\omega}
	\geq \frac{1}{E_m - E_0}\,
	\arccos\!\bigg(
	\frac{\boldr \cdot \boldr_\tau - z^2}{2\wp - 1 - z^2}
	\bigg).
	\label{eq: qubit time}
\end{equation}

Assume that the fidelity between $\rho$ and $\rho_\tau$ is $\delta$. H{\"u}bner's formula~\cite{Hu1992} for the fidelity between qubit states gives
\begin{equation}
    \delta
    = \tr(\rho\rho_\tau) + 2\sqrt{\det\rho\,\det\rho_\tau}
    = \frac{1}{2} (3 + \boldr\cdot\boldr_\tau - 2\wp).  
    \label{eq: qubit fidelity}
\end{equation}
Combining Eqs.~\eqref{eq: the z-coordinate},~\eqref{eq: qubit time}, and~\eqref{eq: qubit fidelity}, we obtain
\begin{equation}
    \tau (E - E_0) 
    \geq \bigg(\frac{1 + z}{2}\bigg)\! \arccos\!\bigg( \frac{2\delta + 2\wp - 3 - z^2}{2\wp - 1 - z^2} \bigg).
    \label{eq: upper bound}
\end{equation}

We now define
\begin{equation}
    \alpha(\delta, \wp)
    = \min\big{\{} f_{\delta, \wp}(z) : z^2 \leq \delta + 2\wp - 2
    \big{\}},
\end{equation}
where the objective function is given by the right-hand side of Eq.~\eqref{eq: upper bound}, or equivalently,
\begin{equation}
    f_{\delta, \wp}(z)
    = (1+z)\arcsin\sqrt{\frac{1-\delta}{2\wp-1-z^2}}.
\end{equation}
For a qubit state of purity $\wp$, any evolution between two states with fidelity $\delta$ in time $\tau$ must satisfy
\begin{equation}
    \tau \geq \frac{\alpha(\delta, \wp)}{E - E_0}.
    \label{eq: qubit QSL}
\end{equation}
In the pure-state limit $\wp = 1$, the qubit speed limit~\eqref{eq: qubit QSL} reduces to the Margolus--Levitin bound~\eqref{eq: the QSL}.

Let $z_{\delta, \wp}$ denote the value of $z$ at which the objective function $f_{\delta, \wp}$ attains its minimum. Suppose that the Bloch vector of a mixed state $\rho$ of purity $\wp$ has $z$-component $z_{\delta, \wp}$, and let $\tau$ be the earliest time at which the fidelity between $\rho$ and $\rho_\tau$ equals $\delta$, so that equalities hold in Eqs.~\eqref{eq: the angle} and~\eqref{eq: qubit time}. Then equality also holds in Eq.~\eqref{eq: upper bound}, yielding
\begin{equation}
    \tau (E - E_0)
    = f_{\delta, \wp}(z_{\delta, \wp})
    = \alpha(\delta, \wp),
\end{equation}
and the bound in Eq.~\eqref{eq: qubit QSL} is saturated.

By adjusting the phases of the energy eigenvectors $\ket{E_0}$ and $\ket{E_m}$, if necessary, we may take the initial Bloch vector to have $x$-component $(2\wp - 1 - z_{\delta, \wp}^2)^{1/2}$ and $y$-component $0$. With this choice of phases, the state assumes the form
\begin{equation}
\begin{split}
    \rho = 
    &\frac{1}{2}(1 - z_{\delta, \wp})\ketbra{E_0}{E_0} \\
    &\,+ \frac{1}{2}\sqrt{2\wp - 1 - z_{\delta, \wp}^2}\,
    \big(\ketbra{E_0}{E_m} + \ketbra{E_m}{E_0}\big) \\
    &\,+ \frac{1}{2}(1 + z_{\delta, \wp})\ketbra{E_m}{E_m}.
\end{split}
\end{equation}
In the pure-state limit we have that $\rho = \ketbra{\psi}{\psi}$, where
\begin{equation}
    \ket{\psi}
    = \sqrt{\frac{1 - z_{\delta, 1}}{2}} \ket{E_0}
    + \sqrt{\frac{1 + z_{\delta, 1}}{2}} \ket{E_m}.
\end{equation}
This establishes the explicit representation in Eq.~\eqref{eq: the pure state}.

\begin{rmk}
    The bound in Eq.~\eqref{eq: qubit time} is reminiscent of the \emph{operational quantum speed limit} for systems with time-independent Hamiltonians introduced in Ref.~\cite{ShLiZhYuLi2020} and further studied in Ref.~\cite{ZhYuLi2023}. In fact, Eq.~\eqref{eq: qubit time} may be regarded as a refinement of the operational quantum speed limit obtained by explicitly accounting for energy conservation. A more detailed analysis of how conserved quantities can be systematically incorporated into the operational framework to obtain progressively stronger lower bounds on the evolution time will be presented elsewhere.
\end{rmk}

\subsection{The dual Margolus--Levitin speed limit}
\label{sec: dual Margolus--Levitin quantum speed limit}
\noindent
Ness \emph{et al.}~\cite{NeAlSa2022} derived a dual version of the Margolus--Levitin quantum speed limit: If a system governed by a time-independent Hamiltonian evolves between two perfectly distinguishable pure states in a time 
$\tau$, then
\begin{equation}
	\tau \geq \frac{\pi/2}{E_m - E},
\end{equation}
where $E_m$ denotes the \emph{largest} populated energy eigenvalue.

This inequality can be extended to an arbitrary overlap $\delta$ between the initial and final states,
\begin{equation}
	\tau \geq \frac{\alpha(\delta)}{E_m - E},
	\label{eq: Ness QSL}
\end{equation}
and it remains valid for mixed states when $\delta$ is taken to be the fidelity between the initial and final density operators.

Equation~\eqref{eq: Ness QSL} follows directly from the Margolus--Levitin bound~\eqref{eq: the QSL} through a time-reversal argument~\cite{HoSo2023a}: If the Hamiltonian $H$ evolves $\rho$ into $\rho_\tau$ in a time $\tau$, then $-H$ evolves $\rho_\tau$ back into $\rho$ in the same time. Let $\delta$ be the fidelity between $\rho$ and $\rho_\tau$. For the reversed Hamiltonian $-H$, the smallest populated eigenvalue is $-E_m$, and the expected energy is $-E$. Applying the Margolus--Levitin bound to this reversed evolution yields
\begin{equation}
    \tau \geq \frac{\alpha(\delta)}{-E - (-E_m)} = \frac{\alpha(\delta)}{E_m - E}.
\end{equation}

If the inequality in Eq.~\eqref{eq: Ness QSL} is saturated, then the support of the initial state is necessarily contained within the eigenspaces of $H$ corresponding to the energies $E_0$ and $E_m$. Moreover, each eigenvector $\ket{\psi}$ of $\rho$ with a nonzero eigenvalue admits a representation of the form
\begin{equation}
	\ket{\psi}
	= \sqrt{\frac{1 + z_\delta}{2}} \ket{E_0}
	+ \sqrt{\frac{1 - z_\delta}{2}} \ket{E_m},
    \label{eq: saturate dual}
\end{equation}
where $\ket{E_0}$ and $\ket{E_m}$ are energy eigenvectors with eigenvalues $E_0$ and $E_m$, respectively. Thus, each such eigenvector represents a pure state that evolves into a state with overlap $\delta$ in time $\tau$, thereby saturating the dual Margolus--Levitin bound~\eqref{eq: Ness QSL}. Moreover, when the initial mixed state is expressed as a convex combination of mutually orthogonal pure states, those associated with nonzero eigenvalues evolve within mutually orthogonal two-dimensional subspaces.

\begin{rmk}
    The sign difference between Eqs.~\eqref{eq: the pure state} and \eqref{eq: saturate dual} implies that the families of states saturating the Margolus--Levitin quantum speed limit and its dual coincide if and only if $z_\delta = 0$, which occurs only for $\delta = 0$. Otherwise, the two families are disjoint. Consequently, the two bounds can be saturated simultaneously only when the fidelity between the initial and final states is zero, in which case saturation of one bound implies saturation of the other.
\end{rmk}

\section*{Acknowledgment}
\noindent
The author gratefully acknowledges Maja Eliasson for inspiring discussions, for contributing to the derivation of the qubit quantum speed limit, and for carefully reviewing early drafts of this paper.

\onecolumngrid

\appendix

\titleformat{\section}[block]{\bfseries\large}{Appendix \Alph{section}:}{0.7em}{}
\titleformat{\subsection}[block]{\bfseries\normalsize}{\Roman{section}.\Alph{subsection}}{0.7em}{}
\titleformat{\subsubsection}[block]{\bfseries\small}{\Roman{section}.\Alph{subsection}.\arabic{subsubsection}}{0.4em}{}

\titlespacing\section{0pt}{1.2em plus 4pt minus 2pt}{0.8em plus 2pt minus 2pt}
\titlespacing\subsection{0pt}{1em plus 4pt minus 2pt}{0.5em plus 2pt minus 2pt}
\titlespacing\subsubsection{0pt}{0.5em plus 4pt minus 2pt}{0.3em plus 2pt minus 2pt}

\section{Proof that $\alpha$ is strictly decreasing}
\label{app: A}
\noindent
In this appendix we show that the function $\alpha$, defined in Eq.~\eqref{eq: the QSL}, is strictly decreasing.

Let $\delta_1 < \delta_2$. Since the objective function $f_{\delta_1}$ is continuous on its compact domain $[-\sqrt{\delta_1}, \sqrt{\delta_1}\,]$, there exists a $z_1$ at which $f_{\delta_1}$ assumes its minimum value. The point $z_1$ also lies within the domain of $f_{\delta_2}$, and because arcsine is strictly increasing, we have that
\begin{equation}
    f_{\delta_1}(z_1) 
    = (1+z_1)\arcsin\sqrt{\frac{1-\delta_1}{1-z_1^2}} 
    > (1+z_1)\arcsin\sqrt{\frac{1-\delta_2}{1-z_1^2}} 
    = f_{\delta_2}(z_1).
\end{equation}
Consequently, $\alpha(\delta_1)
= f_{\delta_1}(z_1)
> \min_{z^2 \le \delta_2} f_{\delta_2}(z)
= \alpha(\delta_2)$, which proves that $\alpha$ is strictly decreasing.

\section{Proof that the objective function has a unique minimum}%
\label{app: B}
\noindent
In this appendix we show that the point at which $f_\delta$ assumes its minimum value is unique. When $\delta = 0$, the domain of $f_\delta$ collapses to the single point $\{0\}$, and the claim is immediate. We therefore assume throughout that $0 < \delta < 1$.

We first observe that the minimum is attained on the interval $[-\sqrt{\delta},0]$, since $f_\delta(-z) < f_\delta(z)$ for all $z>0$. On the open interval $(-\sqrt{\delta},0)$, the derivative of $f_\delta$ takes the form
\begin{equation}
    f_\delta'(z)
    = \arcsin\sqrt{\frac{1-\delta}{1-z^2}}
      + \frac{z\sqrt{1-\delta}}{(1-z)\sqrt{\delta - z^2}}.
\end{equation}
The derivative diverges to $-\infty$ as $z \downarrow -\sqrt{\delta}$, and approaches the positive limit
$\arcsin\sqrt{1-\delta}$ as $z \uparrow 0$. Hence the minimum must occur at a stationary
point in $(-\sqrt{\delta},0)$. To establish uniqueness, it suffices to show that $f_\delta'$ is positive at all of its stationary points, that is, points where the second derivative vanishes, for then the graph of $f_\delta'$ will cross the $z$-axis exactly once.

The second derivative of $f_\delta$ is
\begin{equation}
    f_\delta''(z)
    = \frac{(\delta + 2\delta z - \delta z^2 - 2 z^3)\sqrt{1-\delta}}
    {(1-z)^2(1+z)(\delta - z^2)^{3/2}}.
\end{equation}
It is apparent from this expression that the stationary points of $f_\delta'$ coincide with the zeros of the cubic polynomial $\delta + 2\delta z - \delta z^2 - 2 z^3$. Let $z_* \in (-\sqrt{\delta},0)$ be such a zero. (If no zero exists, $f_\delta'$ has no stationary point in $(-\sqrt{\delta},0)$, and the minimizer of $f_\delta$ is unique.) Then
\begin{equation}
   \delta = \frac{2z_*^3}{1 + 2 z_* - z_*^2}.
   \label{eq: see3}
\end{equation}
Since $0 < \delta < 1$ and $z_* < 0$, the denominator must be negative and the fraction must be strictly less than $1$, which together imply $z_* < -\tfrac{1}{2}$. We use \eqref{eq: see3} to rewrite $f'_\delta(z_*)$ as follows:
\begin{equation}
    f'_\delta(z_*)
    =  \arcsin\sqrt{\frac{1-\delta}{1-z_*^2}} -\sqrt{\frac{1-\delta}{1-z_*^2}}
      \sqrt{1 - \frac{2}{(1 - z_*)^2}}.
\end{equation}
Then, to determine the sign of $f'_\delta(z_*)$ we use that $\arcsin(x) > x$ for $0 < x < 1$ to conclude that
\begin{equation}
    f'_\delta(z_*)
    > \sqrt{\frac{1-\delta}{1-z_*^2}}
      \Bigg(1 - \sqrt{1 - \frac{2}{(1 - z_*)^2}}\,\Bigg).
\end{equation}
Since both factors on the right-hand side are strictly positive, so is $f'_\delta(z_*)$. From this follows that the minimum of $f_\delta$ is unique. We denote the unique minimizer by $z_\delta$.

\section{Proof of the trace-norm properties}\label{app: C}
\noindent
In this appendix we show that $|\tr A|\leq \tr|A|$ for any operator $A$, and that equality holds if and only if $A=e^{i\theta}|A|$ for some real number $\theta$. Since the statement is trivial when $A=0$, we assume $A\ne 0$.

Let $A=V|A|$ be a polar decomposition of $A$, with $V$ unitary. Using the Cauchy--Schwarz inequality for the Hilbert--Schmidt inner product, we obtain
\begin{equation}
    |\tr A|^2
    = \big|\tr(|A|^{1/2} V|A|^{1/2})\big|^2
    \leq  \tr\big(|A|^{1/2} |A|^{1/2}\big) \tr\big(|A|^{1/2} V^\dagger V |A|^{1/2}\big)
    = (\tr|A|)^2.
    \label{eq: C1}
\end{equation}
This proves that $|\tr A|\leq \tr|A|$.

Now assume that equality holds: $|\tr A| = \tr |A|$. Then equality holds in the Cauchy--Schwarz inequality, which implies that the operators $V|A|^{1/2}$ and $|A|^{1/2}$ are linearly dependent. 
Thus, there is a scalar $a$ such that $V|A|^{1/2}=a|A|^{1/2}$. Multiplying this identity with $|A|^{1/2}$ from the right gives $A=a|A|$. Finally, from $|\tr A|=|a|\tr|A|$ and the assumption $|\tr A| = \tr |A|$ it follows that $a$ has unit modulus. Thus, $a=e^{i\theta}$ for some real number $\theta$, and $A=e^{i\theta}|A|$.

\section{Proof that Equation \eqref{eq: trace identity II} implies Equation \eqref{eq: proportional II}}
\label{app: D}
\noindent
In this appendix we show that if Eq.~\eqref{eq: trace identity II} holds, then the compression of $U_\tau$ to the support of $\rho$ is proportional to the orthogonal projection $P$ onto that support. Moreover, we prove that the proportionality constant has modulus $\sqrt{\delta}$, which coincides with the common modulus of the quantities appearing in Eq.~\eqref{eq: trace identity II}. Hence Eq.~\eqref{eq: proportional II} follows.

Let $P_0$ and $P_m$ be the orthogonal projections onto the eigenspaces $\EE_0$ and $\EE_m$ of $H$ corresponding to eigenvalues $E_0$ and $E_m$. Since the support of $\rho$, and thus of $\sqrt{\rho}$, is contained in $\EE_0 \oplus \EE_m$, we have
\begin{subequations}
\begin{align}
	&\rho
	= \sqrt{\rho} P_0 \sqrt{\rho} + \sqrt{\rho} P_m \sqrt{\rho},\\
	&\sqrt{\rho}\, U_\tau \sqrt{\rho} 
	= e^{-i\tau E_0} \sqrt{\rho} P_0 \sqrt{\rho} + e^{-i\tau E_m} \sqrt{\rho} P_m \sqrt{\rho}.
\end{align}
\end{subequations}
From the assumption $\sqrt{\rho} U_\tau \sqrt{\rho}=e^{i\theta}|\sqrt{\rho} U_\tau \sqrt{\rho}|$ it follows that $\sqrt{\rho} U_\tau \sqrt{\rho}$ is normal and therefore commutes with its Hermitian conjugate. A direct computation shows that
\begin{equation}
[\sqrt{\rho}\, U_\tau \sqrt{\rho},\sqrt{\rho}\, U_\tau^\dagger \sqrt{\rho}] = 2i\sin(\tau (E_m-E_0))\,[\sqrt{\rho} P_0 \sqrt{\rho},\sqrt{\rho} P_m \sqrt{\rho}\,].
\end{equation}

Assume that $e^{-i\tau E_0} \neq - e^{-i\tau E_m}$, so that $\sin(\tau(E_m-E_0))\neq 0$. (The case $e^{-i\tau E_0} = e^{-i\tau E_m}$ is trivial since $U_\tau$ then leaves $\rho$ invariant.) Then $\sqrt{\rho} P_0 \sqrt{\rho}$ and $\sqrt{\rho} P_m \sqrt{\rho}$ must commute. Consequently, there exists an orthonormal basis $\ket{\phi_1},\ket{\phi_2},\dots,\ket{\phi_r}$ for the support of $\rho$, consisting of simultaneous eigenvectors of $\sqrt{\rho} P_0 \sqrt{\rho}$ and $\sqrt{\rho} P_m \sqrt{\rho}$, with
\begin{subequations}
\begin{align}
    \sqrt{\rho} P_0 \sqrt{\rho} \ket{\phi_j} &= q_j^{0}\, \ket{\phi_j}, \\
    \sqrt{\rho} P_m \sqrt{\rho} \ket{\phi_j} &= q_j^{m}\, \ket{\phi_j}.
\end{align}
\end{subequations}
The eigenvalues $q_j^{0}$ and $q_j^{m}$ are nonnegative since both operators are positive semidefinite. Moreover, $\ket{\phi_j}$ is an eigenvector of $\rho$ with eigenvalue $p_j = q_j^{0} + q_j^{m}$, which is strictly positive since $\ket{\phi_j}$ belongs to the support of $\rho$. Define $s_j = q_j^{0} / p_j$. Then $0 \le s_j \le 1$, $q_j^{0} = p_j s_j$, and $q_j^{m} = p_j (1 - s_j)$. Introduce the function
\begin{equation}
    q(s)
    = se^{-i\tau E_0} + (1-s)e^{-i\tau E_m},
\end{equation}
which traces the segment between $e^{-i\tau E_0}$ and $e^{-i\tau E_m}$ in the complex plane as $s$ runs over the interval $[0,1]$. Then $\ket{\phi_j}$ is an eigenvector of $\sqrt{\rho}\, U_\tau\, \sqrt{\rho}$ with eigenvalue $p_j q(s_j)$:
\begin{equation}
    \sqrt{\rho}\, U_\tau \sqrt{\rho}\, \ket{\phi_j} 
    = (e^{-i\tau E_0} \sqrt{\rho} P_0 \sqrt{\rho} + e^{-i\tau E_m} \sqrt{\rho} P_m \sqrt{\rho})  \ket{\phi_j}
    = p_jq(s_j) \ket{\phi_j}.
\end{equation}

The condition $\sqrt{\rho} U_\tau \sqrt{\rho} = e^{i\theta}|\sqrt{\rho} U_\tau \sqrt{\rho}|$ requires all these eigenvalues to share the same phase factor $e^{i\theta}$. Thus, the complex numbers $q(s_j)$ must lie on a common ray from the origin. At the same time, each lies on the line segment connecting $e^{-i\tau E_0}$ and $e^{-i\tau E_m}$, which does not pass through the origin under our assumption $e^{-i\tau E_0}\neq - e^{-i\tau E_m}$. Hence, the ray and the segment intersect at a unique point $q$, implying that $q(s_j)=q$ for all $j$. Then
\begin{equation}
    P U_\tau P
    = \sum_{j=1}^r\sum_{k=1}^r \frac{1}{\sqrt{p_jp_k}} \ketbra{\phi_j}{\phi_j} \sqrt{\rho}\, U_\tau \sqrt{\rho} \ketbra{\phi_k}{\phi_k}
    = q \sum_{j=1}^r\sum_{k=1}^r \sqrt{\frac{p_k}{p_j}} \ket{\phi_j}\braket{\phi_j}{\phi_k}\bra{\phi_k}
    = q P.
\end{equation}
Thus, the compression of $U_\tau$ to the support of $\rho$ is proportional to $P$. The modulus of the proportionality constant is
\begin{equation}
    |q|
    = \tr|q\rho| 
    = \tr|\sqrt{\rho}U_\tau\sqrt{\rho}|
    = \sqrt{\delta}.
\end{equation}

It remains to analyze the special case $e^{-i\tau E_0} = -e^{-i\tau E_m}$. Since the purified system described at the beginning of Sec.~\ref{sec: Derivation of necessary conditions} saturates the Margolus--Levitin quantum speed limit, the purification $\ket{w}$ can be written as
\begin{equation}
    \ket{w}=\sqrt{\frac{1-z_\delta}{2}}\ket{E_0} + \sqrt{\frac{1+z_\delta}{2}}\ket{E_m},
\end{equation}
where $\ket{E_0}$ and $\ket{E_m}$ are normalized vectors in $\EE_0 \otimes \HH$ and $\EE_m \otimes \HH$, respectively. Then
\begin{equation}
    \sqrt{\delta}=|\braket{w}{w_\tau}| 
    = \frac{1}{2}\big| (1-z_\delta) e^{-i\tau E_0} + (1+z_\delta)e^{-i\tau E_m}\big|
    = |- e^{-i\tau E_0}z_\delta|=-z_\delta.
\end{equation}
But the analysis in Appendix \ref{app: B} shows that $z_\delta$ lies at an endpoint of the domain of $f_\delta$ only when $\delta = 0$. In this case, $\tr|\sqrt{\rho} U_\tau\sqrt{\rho}|=0$ which forces $|\sqrt{\rho} U_\tau \sqrt{\rho}| = 0$ and therefore $\sqrt{\rho} U_\tau\sqrt{\rho}=0$. It follows that, also in this case, the compression of $U_\tau$ to the support of $\rho$ is proportional to $P$:
\begin{equation}
    P U_\tau P = 0 = \sqrt{\delta} e^{i\theta} P.
\end{equation}
\end{document}